\documentclass[twocolumn,amsmath,amssymb,nofootinbib,prl,superscriptaddress]{revtex4-1}
\pdfoutput=1

\usepackage{graphicx}
\usepackage{dcolumn}
\usepackage{bm,psfrag}
\usepackage[usenames,dvipsnames]{color}

\def\cut#1{{\textcolor{white}{}}}

\newcommand{\be}{\begin{equation}}
\newcommand{\ee}{\end{equation}}
\newcommand{\bea}{\begin{eqnarray}}
\newcommand{\eea}{\end{eqnarray}}
\newcommand{\ba}{\begin{eqnarray}}
\newcommand{\ea}{\end{eqnarray}}

\newcommand{\beq}{\begin{equation}}
\newcommand{\eeq}{\end{equation}}
\newcommand{\beqa}{\begin{eqnarray}}
\newcommand{\eeqa}{\end{eqnarray}}
\newcommand{\beqar}{\begin{eqnarray*}}
\newcommand{\eeqar}{\end{eqnarray*}}

\def\t6 {T_\mt{D6}}

\newcommand{\mt}[1]{\textrm{\tiny #1}}

\def\cale         {{\cal E}}

\def\ee           {{\rm e}}

\def\sqr#1#2{{\vcenter{\vbox{\hrule height.#2pt
 \hbox{\vrule width.#2pt height#1pt \kern#1pt
 \vrule width.#2pt}\hrule height.#2pt}}}}

\def\ee{\cale}

\def\aa1{\phi}
\def\cc1{\psi}

\usepackage[bookmarks=false]{hyperref} 
\hypersetup{pdfstartview=FitH,pdfhighlight=/O,colorlinks=false}

\begin{document}

\title{Reply to ``Comment on two-mode stability islands around AdS''}

\author{Alex Buchel}
\email{abuchel@perimeterinstitute.ca}
\affiliation{Department of Applied Mathematics, Department of Physics and Astronomy, University of Western
Ontario, London, Ontario N6A 5B7, Canada}
\affiliation{Perimeter Institute for Theoretical Physics, Waterloo, Ontario N2L 2Y5,
Canada}
\author{Stephen R. Green}
\email{sgreen@perimeterinstitute.ca}
\affiliation{Perimeter Institute for Theoretical Physics, Waterloo, Ontario N2L 2Y5,
Canada}
\author{Luis Lehner}
\email{llehner@perimeterinstitute.ca}
\affiliation{Perimeter Institute for Theoretical Physics, Waterloo, Ontario N2L 2Y5,
Canada}
\author{Steven L. Liebling}
\email{steve.liebling@liu.edu}
\affiliation{Department of Physics, Long Island University, Brookville, NY 11548, U.S.A}


\maketitle

In a recent Comment~\cite{Bizon:2014bya}, Bizo\'n and Rostworowski
present a few criticisms of our recent
Letter~\cite{Balasubramanian:2014cja}.  In particular, they present
three arguments: (1)~that their own evolutions of the two-mode initial
data we had studied collapse to a black hole around a time
$t\approx 1080$ whereas we found no collapse for times up to roughly
$t=1500$, (2)~that our two-timescale framework~(TTF) ``\emph{cannot be
  even used to infer stability},'' and (3)~that our ``\emph{claims
  about stability islands}'' may not be correct.  \cut{We address each of
these in what follows. We conclude with a discussion of the
implications of their Comment. }

\begin{figure}[h]
\begin{center}
\includegraphics[trim=0.5cm 3cm 0cm 3cm, width=3.2in,clip]{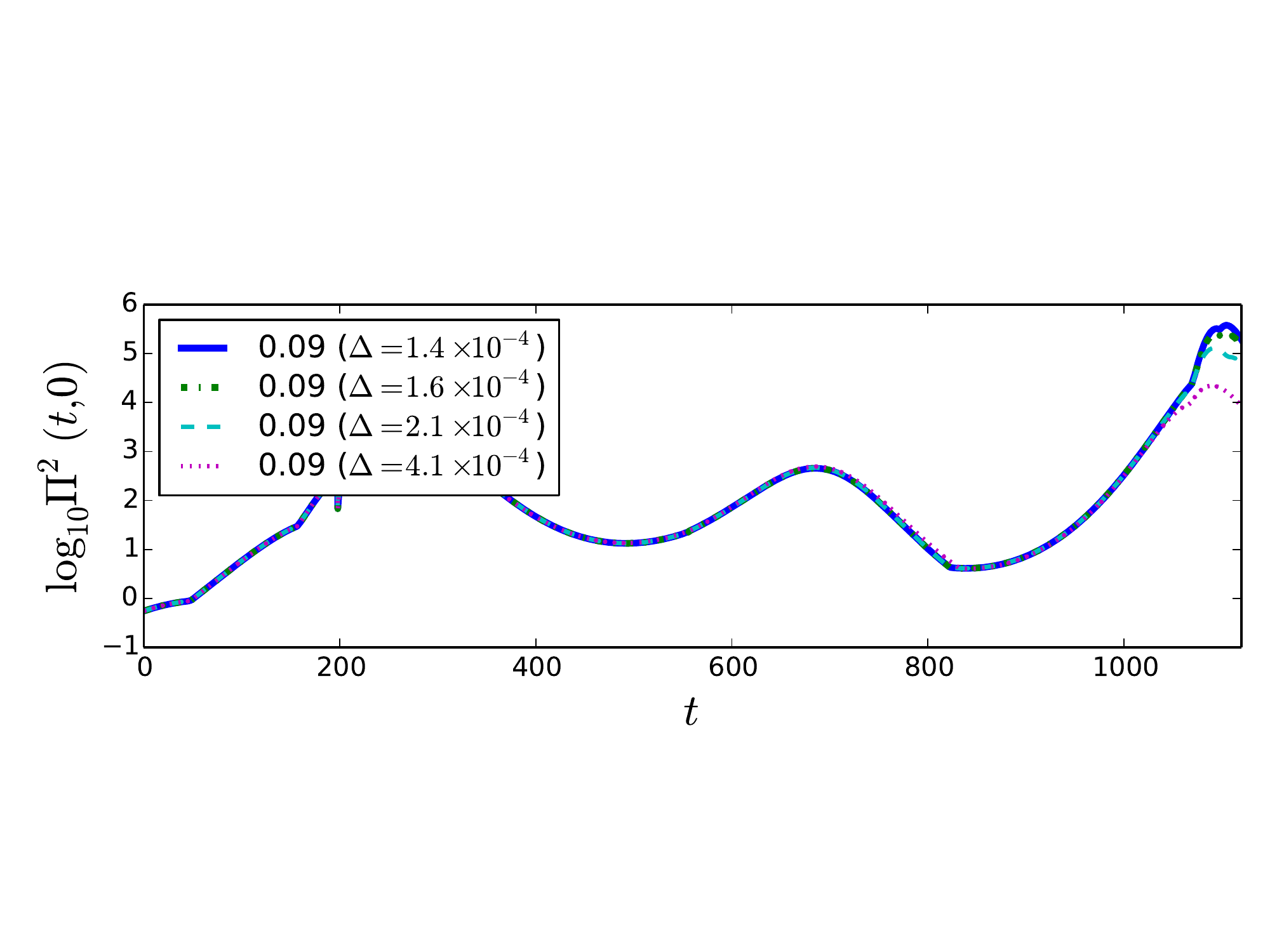}
\end{center}
\caption{\label{fig:1} \cut{(top)} The behavior of the scalar field at the origin 
  for the evolution of equal-energy, two-mode initial data for $\epsilon=0.09$ with increasing resolution.
  \cut{ (bottom) The rescaled behavior for  both $\epsilon=0.08$ and $\epsilon=0.09$ at the same resolution,
    $\Delta=1.6 \times 10^{-4}$.}
  }
 \label{fig:one}
\end{figure}

\cut{In response to their work,} We have studied evolutions of this initial
data with resolutions higher than we had originally. As displayed in
Fig.~\ref{fig:1}, we do find that higher resolutions display higher
concentrations of energy, but we nevertheless have still not observed
collapse to a black hole.  We note that our code demonstrates
convergence to a unique solution even at the late times in question
(i.e., around $t\approx 1080$). We also emphasize that we have
demonstrated that the solution to which our numerical results converge
is in fact a solution of the scalar anti--de~Sitter~(AdS) system we
seek to solve. This latter property (\emph{consistency}) is
demonstrated with tests that the constraint residuals and mass loss
converge to zero (see our Supplementary Material).  And so, while
it is possible that the continuum solution does in fact collapse
around $t\approx 1080$ (and indeed, the work of \cite{nils} does see collapse in a code for which both convergence and consistency are verified), we cannot confirm this
(in time for the publication of~\cite{Bizon:2014bya})
and hence their claim (1).

However even if collapse does take place at a time $t\approx 1080$, the main claims of
our paper still stand. We could not know if collapse occurred after the time which
we ran our code, and so collapse for some time soon after $t=1500$ was always a possibility.
And so if instead collapse occurs at $t\approx 1080$, there is no change
to our claims. Within that time one still finds both direct and inverse cascades
which is now confirmed by the Comment.

Another important point is that, were we to decrease the initial amplitude $\epsilon$,
we would certainly observe any possible collapse pushed to times later than $t\approx 1080$.
\cut{Indeed, in Fig.~\ref{fig:1} we also show $\epsilon=0.08$, properly rescaled, along with the original
value $\epsilon=0.09$.}

Regarding their claims (2) and (3), there seems to be some misunderstanding of what
we tried to communicate in our Letter. Determining whether some scalar perturbation
of AdS with arbitrarily small amplitude $\epsilon$ collapses to a black hole is not
possible with numerical evolutions which require some finite $\epsilon$, finite 
resolution, and finite evolution time $t$.

More particularly regarding their claim (2), we agree that the TTF
cannot be used to infer stability beyond times
$\propto\epsilon^{-2}$. Indeed, we stressed that we have carried our
TTF analysis to $O(\epsilon^3)$ so that one can only trust predictions
within times scaling as $\epsilon^{-2}$. However, for shorter times,
we disagree with the claims of the Comment that our truncation to
$j_{\text{max}}=47$ ``{\em does not suffice to capture the dynamics of the
turbulent cascade}'' for 2-mode initial data. As is clear from Fig.~4
of~\cite{Balasubramanian:2014cja}, the vast majority of the energy
remains in the lowest modes of the system during evolution, and the
dynamics of these modes (in particular, the recurrence times) are very
well-captured by the TTF evolution with $j_{\text{max}}=47$. We have
clearly stated the limitations of our mode-truncation in the
Supplementary Material; in particular, we explained that this leads to the discrepancy
with the full numerical GR curve in Fig.~3
of~\cite{Balasubramanian:2014cja} because the high-$j$ modes are
highly peaked about the origin (despite carrying very small amounts of
energy). We note that the TTF has been verified in Ref.~\cite{Craps:2014jwa}
and even the authors of the Comment have now used the method~\cite{Bizon:2015pfa}.

\cut{
Beyond the practical aspect of generating approximate solutions to the
fully nonlinear AdS-scalar system, we have primarily introduced TTF to
study the energy exchanges as a result of resonant self-interaction of
the field.  The TTF led us to the discovery of a large class of
(apparently stable) quasi-periodic solutions.  It also led us to
postulate a connection between AdS stability and the
Fermi-Pasta-Ulam-Tsingou problem~\cite{Fermi:1955:SNP}.  Furthermore,
Ref.~\cite{Craps:2014jwa} and our work, Ref.~\cite{Buchel:2014xwa},
uncovered the existence of two extra conserved quantities and a
possible hidden symmetry~\cite{Evnin:2015gma} using the TTF formalism.
Finally, we confirmed directly the existence of {\em inverse energy
  cascades} in addition to the direct cascades reported
in~\cite{Bizon:2011gg} and further illustrate how this must happen due
to such conserved quantities~\cite{Buchel:2014xwa}.
}

With regard to claim (3), we did not use the phrase ``two-mode
stability islands'' in our Letter.  In later
work~\cite{Buchel:2014xwa} we referred to ``stability islands,'' but
in that case we were referring to quasi-periodic solutions.

\cut{
From a physical standpoint, the key questions still stand. What types of perturbations
of AdS will lead to collapse? Which ones do not? What physical mechanism governs the
recurrence time observed? Answering these questions is what we find most important, as they would
have profound consequences on our understanding of CFTs through the AdS/CFT correspondence
and/or even help shed further light on such correspondence. Whether a very specific choice
of initial configuration has a particular behavior is certainly of mathematical interest but does
little to shed light on the question of thermalization in AdS. 
Our work, and the introduction of the TTF
formalism which captures key dynamical features was always aimed at this broader set of questions.
}

Their Comment provides independent confirmation
of our main claims: (i)~the presence of both direct and
inverse energy cascades (a fact that was missed in the original perturbative analysis), and (ii)~the validity of the TTF.

{\bf Acknowledgments:} This work was supported by the NSF
under grant PHY-1308621~(LIU), NASA under grant
NNX13AH01G, NSERC through a Discovery Grant (to A.B. and L.L.) and
CIFAR (to L.L.).  Research at Perimeter Institute is supported through
Industry Canada and by the Province of Ontario through the Ministry of
Research \& Innovation.

\bibliography{references}

\end{document}